\documentclass[aps,pre,twocolumn]{revtex4}
\usepackage{graphicx}
\usepackage{amsmath}
\usepackage{amsfonts}
\usepackage{braket}
\usepackage{hyperref}
\usepackage{subfigure}
\usepackage[usenames,dvipsnames]{color}
\usepackage{graphicx}
\usepackage{subfigure}
\usepackage[normalem]{ulem}
\usepackage{soul}

\newcommand{\ct}{\cite}
\newcommand{\la}{\lambda}
\newcommand{\bi}{\bibitem}
\newcommand{\be}{\begin{equation}}
\newcommand{\ee}{\end{equation}}
\newcommand{\ba}{\begin{eqnarray}}
\newcommand{\ea}{\end{eqnarray}}

\newcommand{\non}{\nonumber}
\begin{document}
\title{Probing the role of long-range interactions in the dynamics of a long-range Kitaev Chain}
\author{Anirban Dutta and Amit Dutta \\
Department of Physics, Indian Institute of Technology, Kanpur-208016, India}
\begin{abstract}
We study the role of long-range interactions ({more precisely, the long-range superconducting gap term}) on the non-equilibrium dynamics considering a long-range {}{$p$-wave}
superconducting chain in which superconducting term decays with distance between two sites in a power-law fashion
characterised by an exponent $\alpha$. We show that the Kibble-Zurek scaling exponent, dictating the power-law decay of the defect density in the final state reached following a slow {}{(in comparison to the time-scale associated with the minimum gap in the spectrum of the Hamiltonian)} quenching of the chemical potential ($\mu$) across a quantum critical point, depends non-trivially
on the exponent $\alpha$ as long as $\alpha <2$; on the other hand, for $\alpha >2$, one finds that the exponent saturates to the  corresponding
well-know value of $1/2$  expected for the short-range model. Furthermore, studying the dynamical quantum phase transitions manifested in the non-analyticities in the rate function of the return possibility ($I(t)$) in subsequent
temporal evolution following a sudden change in $\mu$, we show the existence of a new region; in this region, we find three instants of cusp singularities in $I(t)$ associated with
a single sector of Fisher zeros. Notably, the width of this region shrinks as $\alpha$ increases and vanishes in the limit  $\alpha \to 2$ {indicating that  this special region is
an artefact of long-range nature of the Hamiltonian}.

\end{abstract}
\maketitle
\section{Introduction}
  {Remarkable experimental advancement in closed quantum systems of ultracold atoms trapped in optical lattices \ct{bloch08,lewenstein12} and corresponding studies of non-equilibrium dynamics \ct{greiner02,kinoshita06,gring12,trotzky12,cheneau12,schreiber15} of many-body systems have triggered in a plethora of  theoretical
studies.}  At the same time,   numerous experimental studies
of light-induced systems have resulted in the possibility of  non-equilibrium  superconductivity and emergent topological systems \ct{fausti11,rechtsman13}. In parallel,
the theoretical studies investigate  the growth of entanglement entropy following a quench \ct{calabrese06}, thermalization \ct{rigol08}, light-induced topological 
matters \ct{oka09,kitagawa10,lindner11}, dynamics of topologically ordered systems \ct{bermudez09,patel13,thakurathi13}, periodically driven closed
quantum systems \ct{mukherjee09,das10,Russomanno_PRL12,sharma14,bukov16,anirban15,arnab16} and many body localization \ct{pal10,nandkishore15}.
{(For review articles, we refer to \ct{dziarmaga10,polkovnikov11,dutta15,eisert15,alessio16,jstat}. }

  {Very recent  experimental realisation of long-range interacting quantum models with tunable long-range interactions (or long-range pairing term) \cite{expt_longrange} has  revived the interest in studying the  non-equilibrium dynamics of quantum models with infinite-range interactions with interaction strength between two sites separated by a distance $r$  falling off in a power-law fashion as $1/r^\alpha$  \cite{vodola,vodola1,vodola2,regemortel,silva_longrange,fey2016,spin_model_dpt,kzml,dellanna17,unpub,halimeh17,halimeh171,homrighausen17}. Historically, the study of a power-law interacting ferromagnetic Ising chain {}{has a long histrory} \ct{ruelles68,dyson69,dysonn69,kac69}; Thouless \ct{thouless69} studied at length in the context
of the possibility of  long-range order in a one-dimensional system; it was  argued that  a finite critical temperature can not exist for $\alpha>2$. This was later confirmed through renormalisation group calculations
\ct{fisher72}. The most interesting case turns out to be the inverse-sqaure interacting Ising chain  ($\alpha=2$); this,
on the one hand, mimics Kondo effect in metals \ct{yuval71}, on the other,  the transition from the ordered phase to the disordered phase is of  Kozterlitz-Thouless
nature mediated by logarithmically interacting kinks and anti-kinks \ct{kosterlitz76,cardy81,bhattacharjee81,bhattacharjee82,imbrie88,luijten01}. The quantum phase transition (QPT)\cite{sachdev96,suzuki13} in the
corresponding quantum version of the model has also been investigated where as well the situation $\alpha=2$, exhibits a unique behaviour \ct{dutta01}.}

  {While a quantum Ising model with power-law interactions is non-integrable \ct{dutta01,silva_longrange},  recently a generalised version of  $p$-wave superconducting chain \ct{kitaev01} of fermions  with a long-range
{super-conducting pairing/gap term}  has been proposed \cite{vodola,vodola1,vodola2,regemortel,unpub}. The advantage of using this  $p$-wave chain is  the integrability and furthermore, the $2 \times 2$ structure corresponding to each momentum value in spite of the power-law interacting super-conducting term. What is noteworthy, that even though in the short-range limit ($\alpha \to \infty$), the Kitaev chain can be mapped
to the spin-1/2 XY chain through the Jordan-Wigner transformation,  {where the superconducting term represents the spin-spin interaction,} this is not true for the long-range Kitaev chain; the latter, in fact, can not be mapped to a corresponding spin model.}

\ {In this work, we study the effect of long range interactions on the non-equilibrium dynamics of   quantum models exploiting the integrability of the long-range Kitaev chain. In particular, we probe the  Kibble-Zurek (KZ)\cite{Kibble76,Zurek96,Zurek05,ZDZ05,Damski05,Polkovnikov05,damski_zurek06,divakaran08,mukherjee07,sen08,shreyoshi08,anirban16} scaling of the defect density (or the residual energy) when the chain is linearly ramped across one of the quantum critical points (QCPs) and examine the dependence of KZ scaling exponent on the parameter $\alpha$. Furthermore, we subject the chain to a sudden quench across a QCP and analyse the dynamical quantum phase transitions (DQPTs)\cite{heyl13,karrasch13,heyl14,kriel14,divakaran16,pollmann10} (manifested in the non-analyticities in the logarithm of the rate function 
of return probability),  in the sub-sequent temporal evolution with the final time independent Hamiltonian. We show that new features emerge as a consequence of the long-range interactions
which disappear in the short-range limit. It is note-worthy, that the DQPTs \ct{spin_model_dpt,halimeh171,homrighausen17} and Kibble-Zurek scaling \ct{kzml} have been studied for the long-range interacting quantum Ising chains and in both the situations, long-range interactions have been found to play a dominant role, at least for the smaller values of $\alpha$.}

  {The paper is organised in the following fashion: in Sec. \ref{sec_model}, we review the long-range interacting Kitaev chain and illustrate its quantum critical behaviour. Following a brief discussion on the KZ scaling, we analyze the same for the defect density generated following a slow ramp- ing across the QCP in Sec. \ref{sec_KZ}.  In Sec. \ref{sec_dqpt}, on the other hand, we shall consider a sudden quench and analyse the new features those emerge as a consequence of long-range interactions.}

\section{Long-range Kitaev model}
\label{sec_model}

Let us first briefly review the long-range Kitaev chain of fermionic particles residing on a one-dimensional lattice \ct{vodola,vodola1,vodola2}. 
Denoting fermionic annihilation(creation) operators as $c_i (c_i^{\dagger})$, the Hamiltonian can be written as
\begin{eqnarray}
H&=&-J\sum_{i}\left(c_i^{\dagger}c_{i+1}+h.c.\right)-\mu\sum_i\left(n_i-{{1}\over{2}}\right)\nonumber\\
&&+{{\Delta}\over{2}}\sum_{i,l}d_l^{-\alpha}\left(c_i c_{i+l}+c_{i+l}^{\dagger}c_{i}^{\dagger}\right)
\label{ham}
\end{eqnarray}
where $ n_i = c_i^{\dagger} c_i$ is the  number operator for site $i$. Here, $J$ and $\mu$  denote the hopping strength of the fermionic particles between adjacent lattice sites and the onsite chemical potential, respectively, while $\Delta$ is the strength of the superconducting pairing term that decays with distance $l$ in a power law fashion characterized by exponent $\alpha$. {In the limit of $\alpha\rightarrow\infty$, the model reduces to spinless $p$-wave
superconductor Hamiltonian whose topological properties were unravelled by Kitaev \cite{kitaev01}: In an open chain the model Hamiltonian has zero energy Majorana modes at the edges  (and hence the chain is often referred to as a Majorana chain/wire). In this paper, we shall restrict 
our  attention to  a  one-dimensional lattice with $L$ sites  {in a closed ring geometry} (and hence topological properties  of the model will not be exploited}). 
 The term $d_l$ in \eqref{ham} measures the {\it effective} distance between two sites {on the ring} denoted by $i$ and $i+l$ and hence is given by $d_l = \min (l, L - l).$
Furthermore, we shall focus only on the case  $\alpha > 1$. 

{Even in the presence of the long-range
interactions, the Hamiltonian \eqref{ham} is quadratic in fermions and hence}  can exactly be solved
by a Fourier Transformation and followed by Bogoliubov
transformation in terms of fermionic operators in the momentum space $c_i={{1}\over{\sqrt{L}}}\sum_{n=0}^{L-1}e^{-i k_n x_i}c_{k_n}$; {we shall assume an anti-periodic boundary condition} {$(c_i=-c_{i+L})$} so
that  the discrete momenta modes are quantized as  $k_n=({{2\pi}/{L}})(n+{{1}\over{2}})$.
In the Fourier basis, the Hamiltonian (\ref{ham}) can be written in a block diagonal form,
\begin{eqnarray}
H&=&{{1}\over{2}}\sum_{n=0}^{L-1}\Psi_{k_n}^{\dagger}H_{k_n}\Psi_{k_n}
\label{hamf}
\end{eqnarray}
where $\Psi_{k_n}^{\dagger}=\begin{pmatrix}c_{k_n}^{\dagger} & c_{-k_n}\end{pmatrix}$ and $H_{k_n}$ is given by
\begin{equation}
\begin{aligned}
H_{k_n} & = 
\begin{pmatrix}
-(2J \cos k_n+\mu) & i \Delta f_{\alpha}(k_n) \\
-i \Delta f_{\alpha}(k_n) & (2J \cos k_n+\mu)
\end{pmatrix},
\end{aligned}
\end{equation}
where $f_{\alpha}(k_n)=\sum_{l=1}^{L-1}{{\sin (k_n l)}\over{d_l^{\alpha}}}$ {is the Fourier transform of the superconducting gap term}.\\
The Hamiltonian (\ref{hamf}) for each block can be diagonalized by a Bogoliubov transformation
\begin{eqnarray}
\begin{pmatrix}\eta_{k_n} \\ \eta_{-k_n}^{\dagger}\end{pmatrix}=\begin{pmatrix}
\cos \theta_{k_n} & i \sin \theta_{k_n} \\
i \sin \theta_{k_n} & \cos \theta_{k_n}
\end{pmatrix}\begin{pmatrix}c_{k_n} \\ c_{-k_n}^{\dagger}\end{pmatrix};
\end{eqnarray}
where $\tan {\theta_{k_n}}=-{{\Delta f_{\alpha}(k_n)}\over{2J\cos k_n+\mu}}$. The Hamiltonian in the final Bogoliubov basis is given by 
\begin{eqnarray}
H=\sum_{n=0}^{L-1}\lambda_{\alpha}(k_n)\left(\eta_{k_n}^{\dagger}\eta_{k_n}-{{1}\over{2}}\right)
\end{eqnarray}
where $\lambda_{\alpha}(k_n)=\sqrt{(2J\cos k_n+\mu)^2+(\Delta f_{\alpha}(k_n))^2}$ are the eigen energy mode for each $\eta_{k_n}$ fermion. The ground state of the Hamiltonian is given by the vacuum of Bogoliubov $\eta_{k_n}$ fermions and is given by $|GS\rangle=\prod_{n=0}^{L-1}(\cos \theta_{k_n}-i \sin \theta_{k_n} c_{k_n}^{\dagger}c_{-k_n}^{\dagger})|0\rangle$ and $|0\rangle$ is vacuum of $c_{k_n}$ fermions. In the thermodynamic limit $L\rightarrow\infty$, when $k_n$ assume continuous values, the dispersion relation becomes 
\begin{eqnarray}
\lambda_{\alpha}^{\infty}(k)=\sqrt{(2J\cos k+\mu)^2+(\Delta f_{\alpha}^{\infty}(k))^2}
\label{disp}
\end{eqnarray}
  {where $f_{\alpha}^{\infty}(k)= \lim_{L \to \infty} f_{\alpha}(k)={1 \over i}\sum_{l=1}^{\infty}{{e^{ilk}-e^{-ilk}}\over{l^{\alpha}}}=-i\left(\mathbf{Li}_{\alpha}(e^{i k})-\mathbf{Li}_{\alpha}(e^{-i k})\right)$ with  $\mathbf{Li}_{\alpha}(x)=\sum_{l=1}^{\infty}{{x^l}\over{l^{\alpha}}}$ being the Polylogarithmic functions of $x$ that  vanishes in the limit $k\to 0$ and $k\to \pi$. }
Focussing on the situation,  $\alpha>$1, one finds that the spectrum is gapless for the parameter values $\mu=\mp 2J$ for the modes $k=0$ and $\pi$, respectively,
signalling the existence two quantum critical lines in $\mu-J$ plane.  In the subsequent discussion, we shall concentrate on the QCP at $\mu=-2J$ and
analyze the spectrum close to the corresponding critical mode $k=0$. Let us reiterate that  in the short-range  limit  ($\alpha\rightarrow\infty$), the Hamiltonian exactly maps to spin $1/2$ transverse field XY model via Jordan-Wigner transformation. In this case, the critical lines separate the ferromagnetic $(|\mu|<2J)$ and paramagnetic $(|\mu|>2J)$ phases. What is interesting is that the location of the critical lines in the parameter space do not get altered even when
$\alpha$ is finite. For brevity,  we shall henceforth set $J=1$.\\


\begin{figure*}
	\begin{center}
		\includegraphics[width=\columnwidth]{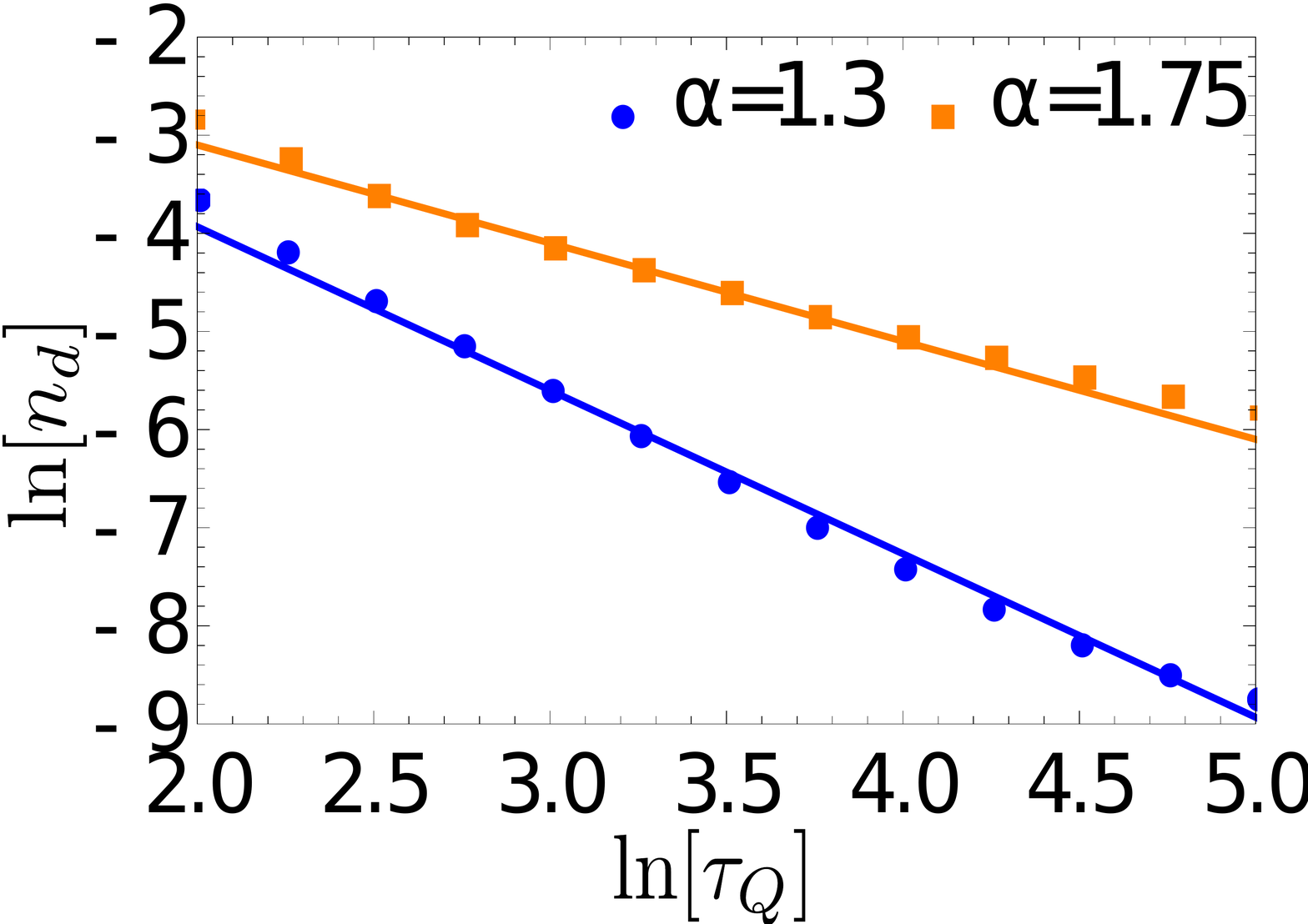}
		\includegraphics[width=\columnwidth]{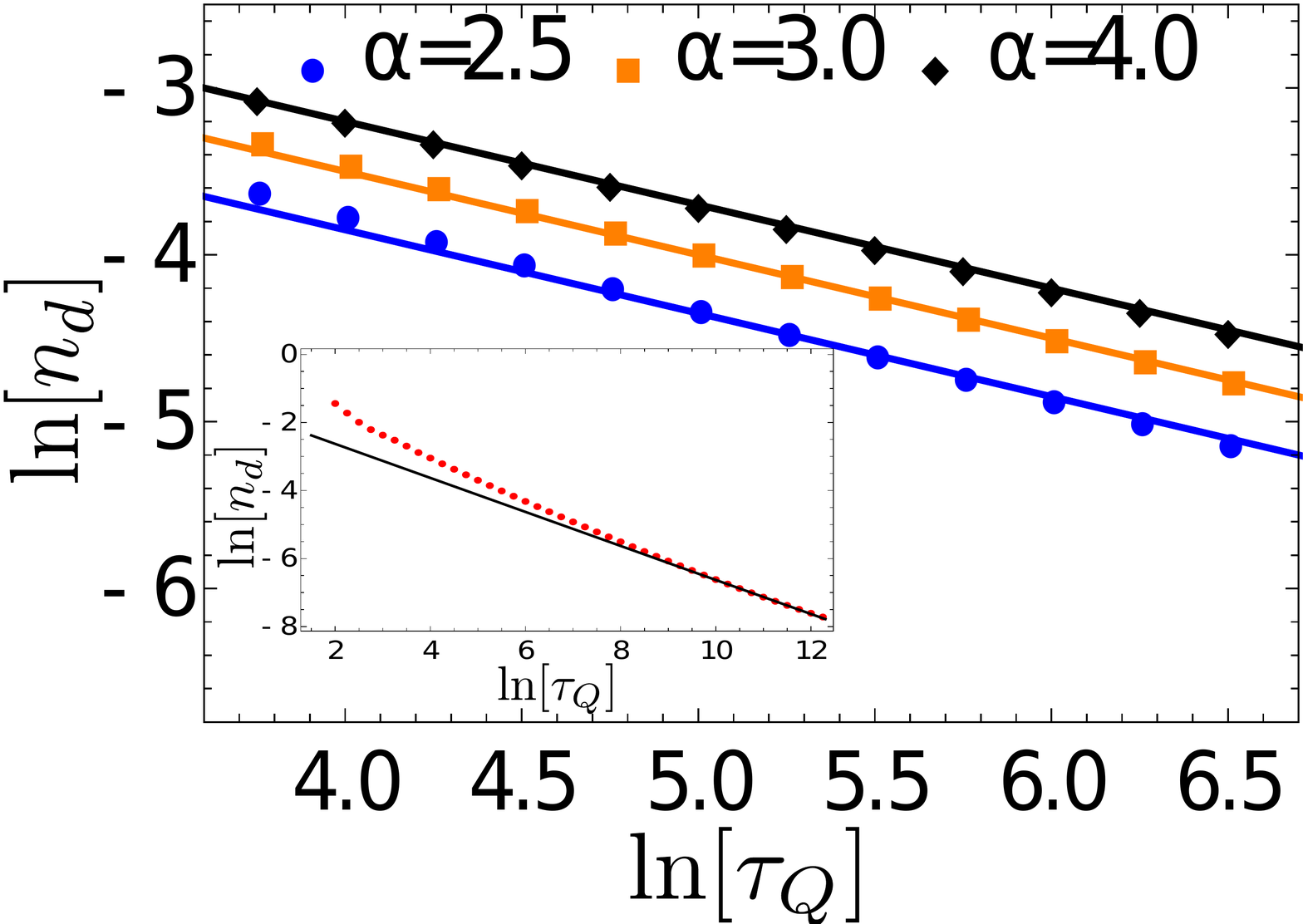}
	\end{center}
\caption{The variation of the defect density $n_d$ with inverse quenching rate $\tau_Q$:
 {\bf (Left)}  The figure shows that the  scaling exponent (determined by the slope of the curve) depends on the  interaction range $\alpha$ for $\alpha <2$. The solid lines show the  theoretically predicted values $1/(2\alpha -2)$  and the dotted lines are from numerical {solution of the differential Eqs. \eqref{lzeq} using Runge-Kutta method with $L=10000$, $\mu_i=-10$ and $\mu_f=0$.}
	{\bf (Right)} The scaling exponent does no longer  change with the range of interaction  for $\alpha \ge 2$, rather gets saturated to  the short range value $1/2$. In the marginal case $\alpha=2$, the exponent reaches the short-range limit of $1/2$ only in the asymptotic limit. }
	
	\label{fig:1}
\end{figure*}

\section{Ramping across the critical point: Kibble-Zurek scaling}
\label{sec_KZ}

  {When a quantum many-body system is quenched across a QCP by changing a parameter (say, linearly as $t/\tau_Q$), the dynamics is necessarily non-adiabatic due to
the diverging relaxation time associated with the QCP; this results in the production of defect (or excess energy) in the final state of the system reached following the quench \cite{Kibble76,Zurek96}.
Remarkably, the density of defect exhibits a universal scaling with the inverse quenching rate $\tau_Q$ and some of the critical exponents associated with the
QCP; this is known as the KZ scaling which has been studied extensively in recent years (For review, see \ct{dziarmaga10,polkovnikov11,dutta15}.) {(A quenching scheme is referred to as a slow
ramping when the  rate of quenching $\tau^{-1}$ is much slower than that associated with the minimum gap in the excitation spectrum of the Hamiltonian; however, close to the QCP, the gap vanishes rendering
the characteristic relaxation time infinite  so that the dynamics is necessarily non-adiabatic however slow may the quenching be.)}

  {To address the question, how does the long-range nature of interaction in \eqref{ham} modifies the KZ scaling relation, we employ
a quenching protocol in which  the chemical potential $\mu$ is tuned as $\mu(t)=-{{t}/{\tau_Q}}$, from a large negative value to a final value $\mu=0$;
in the process the system, initially prepared in the ground state of the initial Hamiltonian, crosses the QCP at $\mu=-2$.
As a result,  the system gets excited in the vicinity of  the critical point and the number of quasi-particles excited in the final state with $\mu=0$ is given by 
\begin{eqnarray}
N=\sum_k \eta_k^{\dagger}\eta_k=\sum_{k}p_k;
\label{noex}
\end{eqnarray}
where  $p_k = \langle\psi_f|\eta_k^{\dagger}\eta_k|\psi_f\rangle$ is the excitation probability in the final state $|\psi_f\rangle$.}

  {Referring to the quasi-momentum mode $k$,  the state of the system at any instant $t$ can be written as  $|\psi_k(t)\rangle=u_k(t)|0\rangle+v_k(t) 
|1\rangle$, where $|0\rangle$ and $|1\rangle$ are the diabatic basis states $(1,0)^T$ and $(0,1)^T$. The corresponding Schr\"{o}dinger equation
is given by:}
{
\begin{eqnarray}
i\dfrac{d }{dt}u_k(t)=-(2J\cos k+\mu(t))u_k(t)+i \Delta f_{\alpha}^{\infty}(k)v_k(t)\nonumber\\
i\dfrac{d }{dt}v_k(t)=-i\Delta f_{\alpha}^{\infty}(k)u_k(t)+(2J \cos k+\mu(t))v_k(t),
\label{scheq}
\end{eqnarray}
with the initial condition $|u_k|^2=1$ and $|v_k|^2=0$. Using the transformation,  $\tau=i\tau_Q\Delta f_{\alpha}^{\infty}(k)\left(t/\tau_Q-2J\cos k\right)$, we can recast  the above Eq.~(\ref{scheq}) to the standard  Landau-Zener (LZ)  form \ct{landau,sei,vitanov}
\ {
\begin{eqnarray}
i\dfrac{d }{d\tau}u_k(t)&=&-(\tau\tilde{\Delta}_k)u_k(t)+v_k(t)\nonumber\\
i\dfrac{d }{d\tau}v_k(t)&=&u_k(t)+(\tau\tilde{\Delta}_k)v_k(t)
\label{lzeq}
\end{eqnarray}
where $\tilde{\Delta}_k=\left(\tau_Q(\Delta f_{\alpha}^{\infty}(k))^2\right)^{-1}$.}}
{For slow passage through the QCP,  the excitation probability  can be calculated using the LZ non-adiabatic transition probability \ct{landau,sei} of that
the system ends in the excited state of the final Hamiltonian $\mu=0$;
\begin{eqnarray}
p_k= e^{-\pi/\tilde{\Delta}_k}\simeq e^{-\pi\tau_Q(\Delta f_{\alpha}^{\infty}(k))^2}.
\label{lztp}
\end{eqnarray}}
\noindent In the large $\tau_Q$ limit, $p_k$ will be significant only for the modes close to the critical mode $k_c=0$ and  hence we use the expansion formula of the polylogarithmic functions in the limit $k\to 0$,
considering three limiting situations and calculate the corresponding scaling of the defect density:

\noindent {Situation I:   $\alpha\not\in\mathbb{Z}$:}\\
In this case, one can use the expansion:

\begin{eqnarray}
(f_{\alpha}^{\infty}(k))^2&=&4\cos^2{{\pi\alpha}\over{2}}\Gamma^2(1-\alpha)k^{2(\alpha-1)}+4\zeta^2(\alpha-1)k^2\nonumber\\&&+8\cos{{\pi\alpha}\over{2}}\zeta(\alpha-1)\Gamma(1-\alpha)k^{\alpha}+o(k^3)\non\\
&=&c_1(\alpha)k^{2(\alpha-1)}+c_2(\alpha)k^{\alpha}+c_3(\alpha)k^{2}+o(k^3)\non\\
\label{polyee1}
\end{eqnarray}
 {where we have used standard Gamma functions($\Gamma$) and Riemann zeta functions($\zeta$) \cite {vodola,functions}  \  {and the coefficients are $c_i(\alpha)$s given by:}
	\begin{eqnarray}
	c_1(\alpha)&=&4\cos^2{{\pi\alpha}\over{2}}\Gamma^2(1-\alpha),\\
	c_2(\alpha)&=&8\cos{{\pi\alpha}\over{2}}\zeta(\alpha-1)\Gamma(1-\alpha),\\
	c_3(\alpha)&=&4\zeta^2(\alpha-1).
	\end{eqnarray}}
	
Using the expression of Eq.(\ref{polyee1}) in  Eq.(\ref{noex}),  the density of quasiparticle excitation in the final state  at the end of the drive  in the thermodynamic limit   can be calculated as
\begin{eqnarray}
n_d&=&N/L={{1}\over{2\pi}}\int_{-\pi}^{\pi}p_k dk \approx {}\int_{-\infty}^{\infty}p_k dk \nonumber\\
&\simeq&{{1}\over{2\pi}}\Big({{1}\over{(\pi c_1(\alpha)\tau_Q)^{{1}\over{2\alpha-2}}}}+{{1}\over{(\pi c_2(\alpha)\tau_Q)^{{1}\over{\alpha}}}}+\nonumber\\&&{{1}\over{(\pi c_3(\alpha)\tau_Q)^{{1}\over{2}}}}\Big),
\label{kzf1}
\end{eqnarray}
where we have extended the range of integration over $k$ from $-\infty$ to $\infty$.

\noindent The expansion formula for polylogarithmic functions for integer $\alpha$ is given by 
\begin{eqnarray}
f_{\alpha}^{\infty}(k)&=&-\frac{i^\alpha k^{\alpha-1}}{(\alpha-1)!}\left[(1+(-1)^\alpha)(\mathbb{H}_{\alpha-1}-\ln{k}-\frac{i\pi}{2})+i\pi\right]\nonumber\\&&-2\sum_{\substack{m\in odd \\ m\neq \alpha-1}}\frac{\zeta(\alpha-m)i^{m+1}}{m!}k^m,
\label{polyee2}
\end{eqnarray}
where $\mathbb{H}_n$ is nth Harmonic number. Using this expansion, we shall probe the following cases:\\

\noindent Situation II: $\alpha$ is an integer $\neq 2$:\\

For any  $\alpha\neq2,\in\mathbb{Z}$ one finds: 
\begin{eqnarray}
(f_{\alpha}^{\infty}(k))^2&=&4(\zeta(\alpha-1))^2 k^2+o(k^3).
\label{pkintk}
\end{eqnarray}

\noindent {Using Eq.~\eqref{pkintk}, one can similarly  calculate the  scaling of the density of quasiparticle excitation in the final state
\begin{eqnarray}
n_d&\simeq&{{1}\over{(4\pi(\zeta(\alpha-1))^2 \tau_Q)^{{1}\over{2}}}}. 
\label{kz2}
\end{eqnarray}

\noindent Situation III: Marginal case $\alpha =2$\\

 The expansion of $(f_{\alpha}^{\infty}(k))^2$ for $\alpha$=2 can be calculated from the expansion Eq.(\ref{polyee2}):
\begin{eqnarray}
(f_{2}^{\infty}(k))^2&=&4(1-\ln k)^2 k^2+o(k^3);
\label{pk2k}
\end{eqnarray}

\noindent We note that for $\alpha$=2,  there is a prominent  logarithmic correction in the leading order to the expression of $p_k$ which leads to sub-leading corrections to the scaling of $n_d \sim 1/\sqrt{\tau_Q}$. However
in the asymptotic limit  of $\tau_Q \to \infty$, the sub-leading corrections drops off  yielding the short-range scaling relation  as shown in the inset of the right panel of  Fig.~\ref{fig:1}.

  {Let us now inspect the scaling relation predicted in Eqs.~\eqref{kzf1} and \eqref{kz2}, recalling that in the $\alpha \to \infty$ limit $n_d \sim \tau^{-1/2}$; interestingly,
we find that $1<\alpha<2$, the KZ scaling exponent   is determined  by the first term in Eq.~\eqref{kzf1} having the slowest decay with $\tau_Q$   in the limit of $\tau_Q \to \infty$. We thus have the KZ scaling exponent $1/(2\alpha -2)$,  which reduces to the short-range value when $\alpha \to 2$.For $\alpha >2$, on the contrary,   scaling exponent  gets saturated to $1/2$.  This establishes that
the case $\alpha=2$ marks the boundary between the long-range and the short-range behavior so far as the KZ scaling is concerned.  Furthermore, in the marginal case $\alpha=2$, there are
non-universal sub-leading corrections which vanish in the limit of $\tau_Q \to \infty$. 
 {To verify the above mentioned analytically predicted scaling
relations for different ranges of $\alpha$, we have numerically solved the differential Eqs. \eqref{lzeq} to calculate the excitation probability and hence the defect density as shown in  Fig. \ref{fig:1}; we find that there is an excellent agreement between the numerical and analytical results.}

\section{Dynamical Phase Transition following a sudden quench}

\label{sec_dqpt}

  {Dynamical Quantum Phase transitions (DQPTs),  introduced by Heyl  $et~ al.$ \cite{heyl13}, in the context of non-analyticities in the temporal evolution of a  quenched  quantum system, is one of the emerging features} of non-equilibrium dynamics of closed quantum systems.   {For a sudden quenching scheme}, the system is prepared in the ground state $|\psi_g\rangle$ of the Hamiltonian $H_i$ corresponding to a parameter $\lambda=\lambda_i$. At time $t=0$,   {the parameter $\lambda$ of the Hamiltonian is suddenly changed from a value $\lambda_i$ to $\lambda_f$ while the state remains frozen  at  $|\psi_f\rangle=|\psi_g\rangle$.  The initial state then has a non-trivial temporal  evolution generated by the time-independent final Hamiltonian $H_f (\la_f$) and one defines }  {the so-called Loschmidt} overlap amplitude(LOA)   {at an instant $t$} as  $G(t)=\langle\psi_f|e^{-i H_f t}|\psi_f\rangle$. Generalising to the complex plane $G(z)=\langle\psi_f|e^{- H_f z}|\psi_f\rangle$,   {where $z=R +it$ with $R$ being the
real part},  we can now define the dynamical free energy density in the thermodynamic limit for a $d$-dimensional system with linear dimension $L$  
\begin{eqnarray}
f(z)=-\lim_{L\rightarrow\infty}{1\over{L^d}}\ln{G(z)}.
\end{eqnarray}
  { The zeros of $G(z)$ (which also indicate non-analyticities in $f(z)$) are so-called  ``Fisher zeros" (FZ)\cite{fisher65,lee52,saarloos84} which  form a line in the complex $z$ plane for $d=1$ in the thermodynamic limit. In case, this line crosses the imaginary  (real time) axis,
which {\it usually} happens for quenches across a QCP, one observe cusp singularities in the rate function of the return probability defined as $I(t)=-\log|G(t)|^2/L$. Sharp non-analyticities of $I(t)$, at those instants of real time, referred to as DQPTs.  Various aspects of DQPTs  for several systems  have been extensively
studied in recent years\cite{heyl15,palami15,divakaran16,heyl16,huang16,puskarov16,zhang16,vajna14,sharma15,vajna15,schmitt15,budich15,andraschko14,canovi14,pollmann10,sharma16,zvyagin17,sei17,fogarty17,utso16,utso17,bhattacharya17,
heyl17}.}\\

\begin{figure*}[]
		\includegraphics[width=0.9\columnwidth]{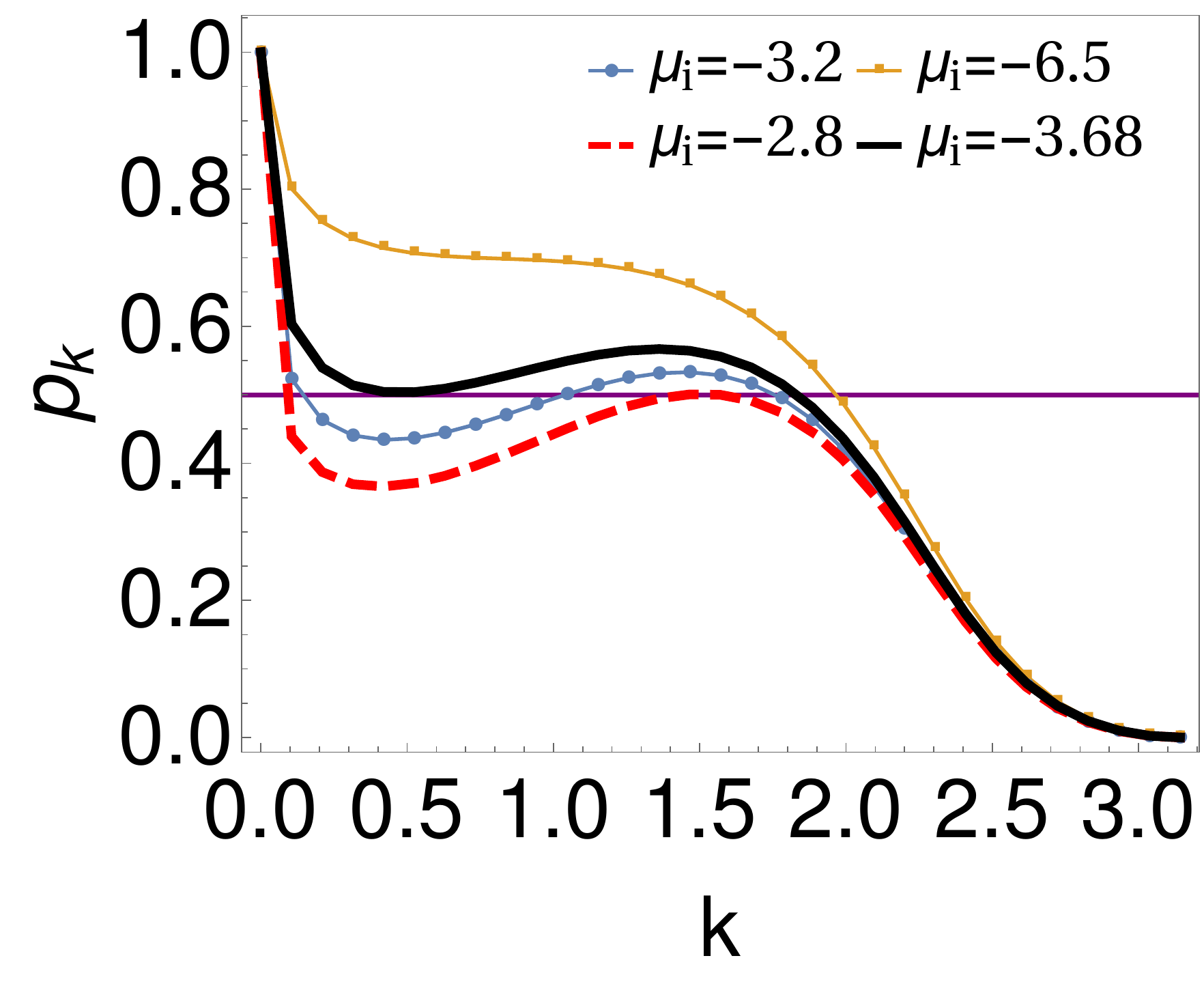}
		\includegraphics[width=0.9\columnwidth]{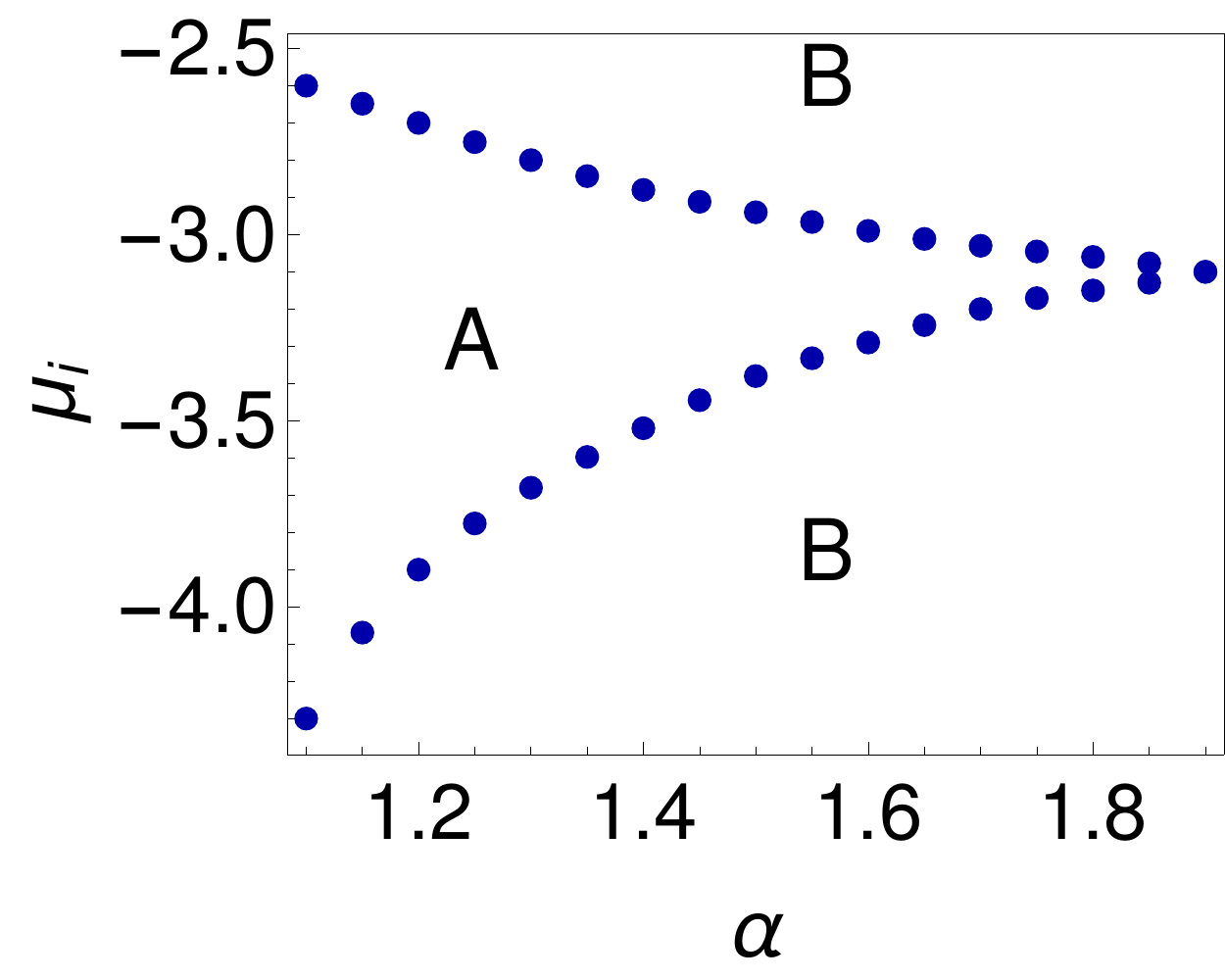}
		\caption{
	{\bf (left)} Plot of the  transition probability   {\eqref{eq_p_k}} as a function of $k$ for different values of $\mu_i$ with $\mu_f=1$;  the long-range interaction parameter  $\alpha$ is fixed to $\alpha=1.3$. We find  two distinct regimes: (i) The transition probability $p_k$ crosses the value 1/2  for three values of $k$  (i.e., yielding three values of  $k_*$) for  $\mu_i=-3.2$ and once for  $\mu_i=-6.5$, while for other two values it crosses the $p_k=1/2$ once and touches at another value of $k$.	{\bf (Right)} Phase boundary separating  the three crossings region (A) from the single crossing regions (B)  is shown in the $\alpha-\mu_i$ plane. On the phase boundary  $p_k$ crosses (and touches) the $p_k=1/2$ line  twice as shown in the left panel. It is noteworthy that the width of region A shrinks as $\alpha$ increase
	and vanishes as $\alpha \to 2$.
	}
	\label{fig:2}
	\end{figure*}
  {In this paper, we shall study the effect of long-range interactions, namely the parameter $\alpha$ on the DQPTs following a  {\it sudden} quench of the parameter  $\mu$ of
the Hamiltonian  from an initial value $\mu_i(< -2)$ to $\mu_f=1$ across a critical point $\mu=-2$ while  the state of the system stays  frozen in the ground state of the initial Hamiltonian. For the mode 
$k$, one can express the ground state for the mode $k$,  $|\psi_{0}^k \rangle$ as,
\begin{eqnarray}
|\psi_0^k\rangle=u_f(k)|1_k^f\rangle+v_f(k)|2_k^f\rangle;
\end{eqnarray}
where $|1_k^f\rangle$ and $|2_k^f \rangle$  denote  the ground(excited) state of the final Hamiltonian $H_k^f=H_k(\mu_f)$ with energy eigenvalues $\epsilon_k^f$ and $-\epsilon_k^f$, respectively, with 
  $|u_f(k)|^2+|v_f(k)|^2=1$; the dynamical free energy  in the complex $z$ plane is then given by  
 $$f_k(z)=-\log\left(|v_f(k)|^2 \exp(\epsilon_{k}^f z)+|u_f(k)|^2 \exp(-\epsilon_{k}^f z)\right)$$}
 Integrating over all the momenta modes we get,
\begin{eqnarray}
&~&f(z)=\nonumber\\
&&\int_0^{\pi}\frac{dk}{2\pi}\log\left(|v_f(k)|^2 \exp(\epsilon_{k}^f z) +|u_f(k)|^2 \exp(-\epsilon_{k}^f z).\right)\nonumber\\
&=&-\int_0^{\pi}\frac{dk}{2\pi}\log\left((1-p_k)\exp(\epsilon_{k}^f z) +p_k\exp(-\epsilon_{k}^f z)\right);\nonumber\\
\label{eq_free_energy}
\end{eqnarray}
The zeros of the dynamical partition function $G(z)$ (i.e., FZs) is then given by
\begin{eqnarray}
z_n(k)={1\over{2 \epsilon_k^f}}\left[\log\left({p_k\over{1-p_k}}\right)+i \pi(2n+1)\right]
\label{fz}
\end{eqnarray}
  {where $n=0,1,2 ...$, $p_k=|v_f(k)|^2$ is the probability of excitation. 
The critical mode $k=0$ is frozen and $p_k=1$ while far away from critical modes $p_k \to 0$. {This implies that as the lattice momentum varies from $0$ to $\pi$, the real values
of FZs  given in Eq.~\eqref{fz} span from $-\infty$ to $\infty$ along the real time axis}. For an intermediate value  $k=k_*$ when  $p_{k_*}=1/2$, the lines of
FZs lie  {right} on the imaginary (real) time axis; the corresponding instants of real time are}
\be
t_n^* =  \frac{\pi} {2\epsilon_{k_*}^f}  \left(2n+ 1 \right).
\label{eq_time}
\ee
  {The rate function of the return probability in this case can be evaluated
exactly in the form \ct{heyl13}}
\ba
I(t) &=&-\frac{\log |G(t)|^2}{L} = 2~ {\rm Re} f(z)\non\\
&=&- \int_{0}^{\pi} \frac{dk}{2\pi} \log\left(1 + 4 p_k (p_k-1) \sin^2 (\epsilon_{k}^f t) \right); \non\\
\label{eq_rate_function}
\ea
  {are non-analytic due to non-analytic contribution for the mode $k_*$ at the real instants $t_n^*$ described by Eq. \eqref{eq_time}.} {The physical
  significance of the mode $k_*$ with $p_{k_*}=1/2$  is that  both the ground and the excited states of the final Hamiltonian for the mode $k_*$ are equally
  populated; (this mode $k_*$, in that sense, is at infinite temperature with respect to the final Hamiltonian). The existence of a  momentum mode $k_*$ ensures the
  existence of DQPTs; if the quenching amplitude is such that there exists no $k_*$, the DQPTs will be absent in the subsequent evolution generated by the final
  Hamiltonian. }

\begin{figure*}[]
	\begin{center}
	\includegraphics[width=\columnwidth]{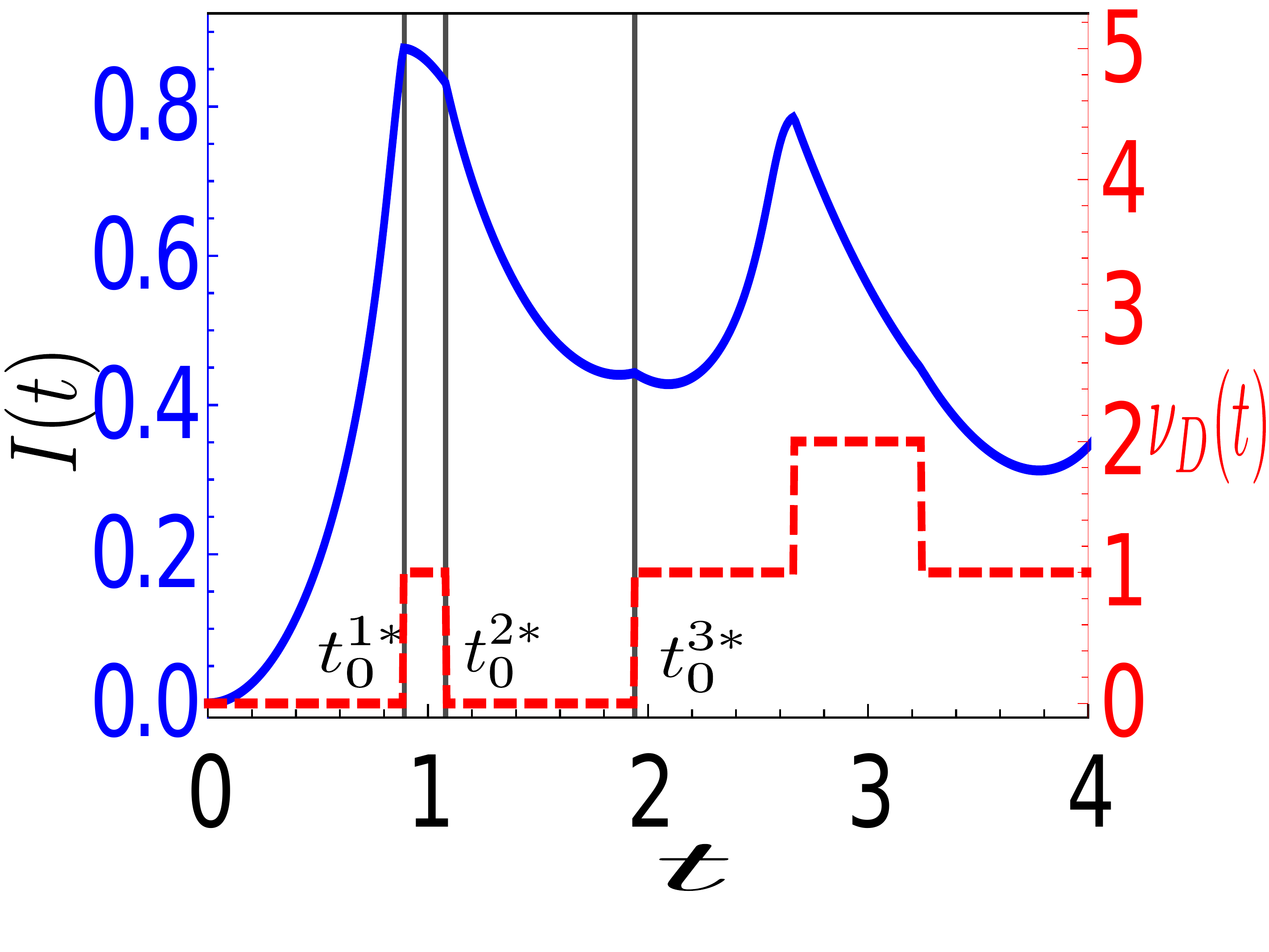}
	\includegraphics[width=\columnwidth]{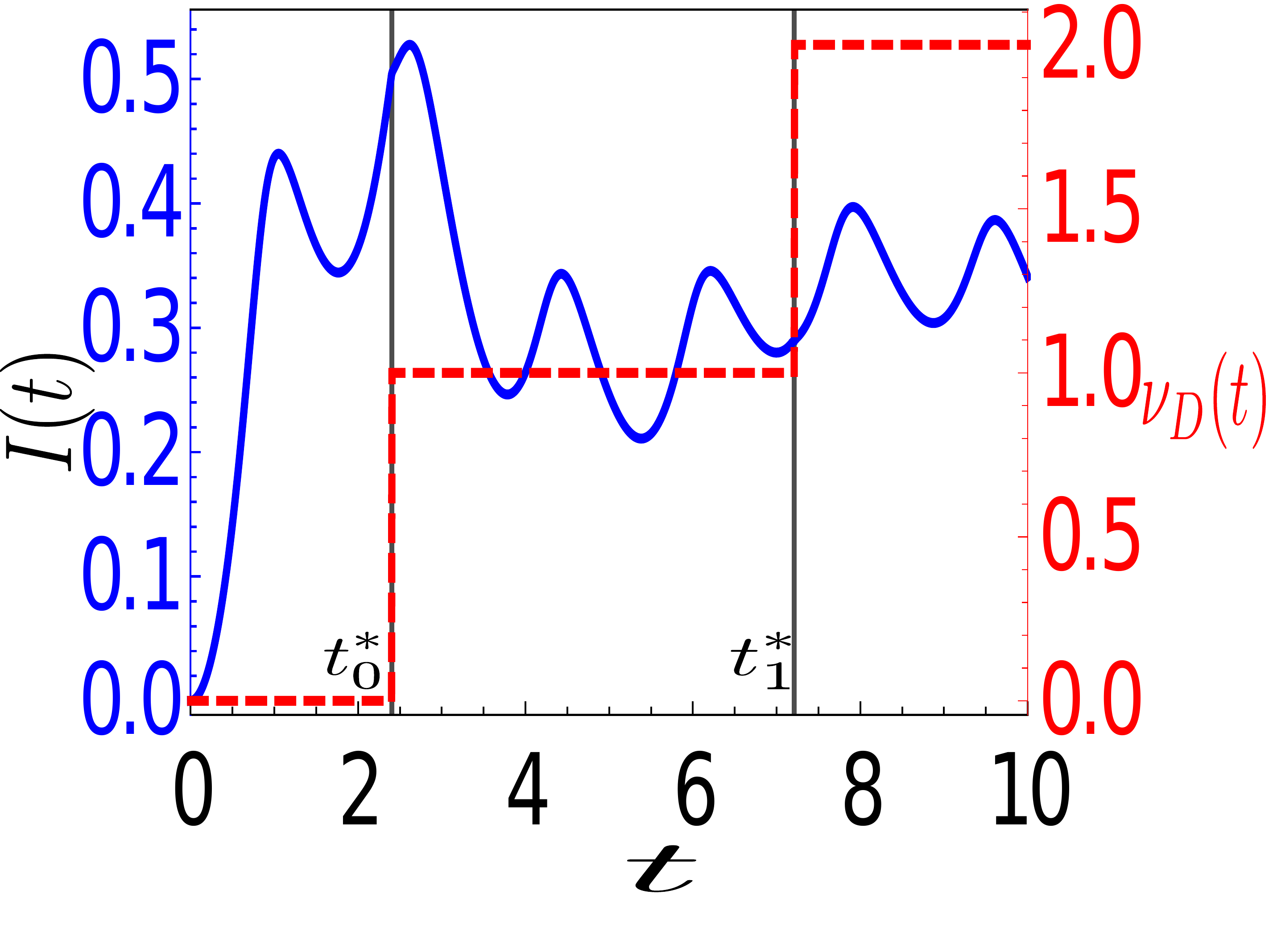}
	\end{center}
	\caption{({\bf left}) The temporal evolution of $\nu_D(t)$ {as a function of $t$} corresponding to the  region A  shown in the right panel of Fig.~\ref{fig:2}. We find
	that corresponding to $n=0$, there are three instants $t_0^{1*}$, $t_0^{2*}$ and $t_0^{3*}$ at which $\nu_D(t)$ makes discontinuous changes by a factor of unity. However,
	whether the DTOP jumps or drops is determined by the slope of the $p_k$ vs $k$ curve at $k_*$ (i.e., the crossing points) as depicted in the left panel of the Fig.~\ref{fig:2}.
	({\bf right}) $\nu_D$ vs $t$ for the region B the right panel of Fig.~\ref{fig:2}, we have only one instant of non-analyticity corresponding to each $n$ and we show jumps in the DTOP corresponding 
	to $n=0$ and $n=1$; notably in this case there is no drop in $\nu_D$ (t) at DQPTs. }
	\label{fig:3}
\end{figure*}

From the   {$2 \times 2$} Hamiltonian  (\ref{hamf}) we find that the ground state and excited states of the initial and final Hamiltonian,{$ |1^{i/f}\rangle$ and  $|2^{i/f}\rangle$, respectively,}  {can be 
decomposed to the form using the {diabatic} basis vector $(1,0)^T$ and $(0,1)^T$ (where $T$ denotes the transpose) in the form: } 
\begin{eqnarray}
|1^{i/f}\rangle=\cos {\frac{\theta_k^{i/f}}{2}} (1,0)^T-\sin {\frac{\theta_k^{i/f}}{2}} (0,1)^T;\nonumber\\
|2^{i/f}\rangle=\sin {\frac{\theta_k^{i/f}}{2}} (1,0)^T+\cos {\frac{\theta_k^{i/f}}{2}} (0,1)^T;
\end{eqnarray}
where $\tan\theta_k=\Delta_k/(\mu+2\cos k)$ \ {and $\Delta_k=\Delta f_{\alpha}(k)$ depends on $\alpha$.} The excitation probability $p_k$ in this notation simplified as,
\begin{eqnarray}
&&p_k=|\langle 1^i|2^f\rangle|^2=\sin^2\left[(\theta_k^i-\theta_k^f)/2\right]\nonumber\\
&=&{\frac{1}{2}}\left[1-{\frac{(\mu_i+2\cos k)(\mu_f+2\cos k)+\Delta_k^2}{\sqrt{((\mu_i+2\cos k)^2+\Delta_k^2)((\mu_f+2\cos k)^2+\Delta_k^2)}}}\right]\nonumber\\
\label{eq_p_k}
\end{eqnarray}
From the above discussion,  one  can conclude the condition for DQPT as there exists definite momentum mode(s) defined as $k=k_*$ for which the $p_{k_*}=1/2$ and for this mode we have the condition $(\mu_i+2\cos k_*)(\mu_f+2\cos k_*)+\Delta_{k_*}^2=0$   {for all values of $\alpha >1$. We note that for $\alpha \to \infty$, the prominent occurrence of DQPTs have already been established \cite
{heyl13,vajna14}.}

 {However, there is an interesting possibility that arises due to the long-range nature of the model;  we illustrate this  below assuming for simplicity $\mu_f=1$.   {Let us refer to the left panel of Fig.~\ref{fig:2}; interestingly, we find
that   there exists a region in the parameter space of   $\mu_i$, where   there are  three  values of $k_*$  (as $k$ ranges from $0$ to $\pi$)  for which $p_{k_*}=1/2$  even though the system is quenched across a single QCP.} For each of the values of $k_*$, we have different values of $t_{n}^*$ for the same value of $n$ as obtained from Eq.~\eqref{eq_time}. Usually, one encounters a single instant of non-analyticity for a given $n$, i.e., 
there is a single $k_*$ as $k$ ranges from $0$ to $\pi$, following a quench across a QCP.}

In the right panel of Fig.~\ref{fig:2},  we  plot a phase diagram in $\mu_i -\alpha$ plane with $\mu_f=1$, \ {separating these two regions: (i) in region A, there are
three $k_*$ values leading to three instants of real time at which DQPTs occur for a given $n$ as shown in the left panel of Fig.~\ref{fig:3}. (ii) On the contrary in region B, there is only $k_*$ as in the short-range case (right panel Fig.~\ref{fig:3}). Finally, on the phase boundary  separating these two regions, there are two values of $k_*$ with $p_{k_*}=1/2$.  What is note worthy
that $\alpha \to 2$, the width of the region A shrinks and vanishes confirming that the occurrence of a region A with three DQPTs for a given $n$ is indeed an artefact of long-range
nature of the Hamiltonian.   

  {Let us finally address the question whether the DQPTs occurring in this model  can be characterised by the  dynamical topological order parameter (DTOP)
introduced in Ref. \cite{budich15}, and how does its behaviour change in the region A as compared to the region B. {We would like to mention that the DTOP can be interpreted as  a non-local order
parameter and  its temporal variation appropriately characterises every instant at which DQPTs occur.  More precisely, it stays fixed to a quantised integer value between two instants of successive DQPTs while
at every DQPT it shows a discontinuous jump of unit magnitude. } Writing}
the LOA  for each $k$ mode   {in the polar form as} $f_k(t)=|r_k(t)|\exp{i \phi_k(t)}$ where $\phi_k(t)$ is given by the formula
\begin{eqnarray}
\phi_k(t)=\tan^{-1}\left\{(|v_f(k)|^2-|u_f(k)|^2)\tan(\epsilon_k^f t)\right\},
\end{eqnarray}
  {we can define the gauge-independent Pancharathnam geometric phase (PGP) in the form $\phi_k^G(t)=\phi_k(t)-\phi_k^{\rm dyn}(t)$, where }
the dynamical phase $\phi_k^{\rm dyn}(t)=-\int_{0}^{t}ds\langle\psi_f^k(s)|H_k^f|\psi_f^k(s)\rangle=(|v_f(k)|^2-|u_f(k)|^2)\epsilon_k^f t$.   {The DTOP ($\nu_D(t))$ is then defined as the winding number  }
\begin{eqnarray}
\nu_D(t)={\frac{1}{2\pi}}\int_{0}^{\pi}\frac{\partial \phi_k^G}{\partial k}
\label{eq_dtop}
\end{eqnarray} 
{The phase $\phi_k^G$ is pinned to zero for the momentum values   $k=0$ and $\pi$,i.e., $\nu_D(t)$ defined in Eq.~\eqref{eq_dtop} stays constant as long as the system does hit a non-analyticity at the critical time when there is
a discontinuous jump \ct{budich15,sharma16} as shown in Fig.~ \ref{fig:3}; whether the jump will be increasing or decreasing ($+1$ or $-1$) will be determined by the slope of the $p_k$, i.e., $\partial p_k/\partial k$ at
the corresponding critical momentum $k_*$. }
\ {Referring to the left panel of  Fig.~\ref{fig:3}, the first three  non-analyticities  in $I(t)$ correspond  to the sector $n=0$  as obtained from Eq. \eqref{eq_time} for three values of $k_*$ as
depicted in the curve corresponding to $\mu_i=-3.2$ in the left panel of the  Fig. \ref{fig:2}. Labelling these  instances,  derived from \eqref{eq_time}, as $t_0^{1*}$, $t_0^{2*}$ and $t_0^{3*}$, respectively, we observe that the DTOP  jumps at $t_0^{1*}$
 and $t_0^{3*}$ while at $t_0^{2*}$ it drops,  in all  the cases  by  a factor of unity. The behaviour of the DTOP, i.e, whether $\nu_D$ would  jump or drop, is determined by the slope the curve $p_k$ vs $k$ at $k_*$  \ct{budich15,sharma16}; following the same curve in Fig.~ \ref{fig:2}, we find  the slope is different for the middle crossing in comparison to the other two crossings which
 results in the change of polarity of the jumps in $\nu_D(t)$ at a DQPTs. On the contrary, for region B , we have only crossing for one value of $n$ and we observe successive
 jumps of the same polarity at every DQPTs as shown in the right panel of Fig.~\ref{fig:3}.  }

 \section{Concluding comments}
 
   {Using the integrability and the $2 \times 2$ nature of a Kitaev chain  with a long-range super-conducting term,  we have studied the effect of long-range interactions on the KZ scaling for slow quenching  and DQPTs following a sudden quenching.  In both the situations,  we have established that the long-range interactions characterised by the parameter
 $\alpha$ plays a non-trivial role. In the long-range interacting case $1<\alpha <2$, we show that the KZ scaling exponent dictating the power-law decay of the defect density ($n_d$)
 with the inverse rate of quenching ($\tau_Q$) depends non-trivially on $\alpha$ and is given by $1/(2\alpha-2)$. In the limit of $\alpha \geq 2$, i.e., in the short-range limit, the exponent saturates to
 the short-range value $1/2$. Therefore, the value $\alpha=2$ marks the boundary between the long-range and the short-range behaviour of the scaling of $n_d$. It is note-worthy that
the scaling of $n_d$ in  the marginal case $\alpha=2$ has non-universal sub-leading corrections which vanish in the limit $\tau_Q \to \infty$. }
{Let us recall that the importance of the case $\alpha=2$ for a classical one dimensional Ising model emerges from a renormalisation group  calculations (and the transition at $\alpha=2$ is of Kosterlitz Thouless nature). However, the present work deals with a quantum model in a non-equilibrium situation where the importance of the $\alpha=2$ scenario emerges from the fact that the scaling behavior of the excitation spectrum at the quantum critical point (with the momentum) changes precisely at $\alpha=2$. Consequently the case $\alpha=2$ points to a crossover from the long-range to the short-range behavior. Whether this can be put in a generic renormalization scenario for a non-equilibrium situation is indeed a pertinent question for future research.}

  {For DQPTs, we observe an interesting three crossing region where there exist three instants of non-analyticities in $I(t)$ for
a given $n$.  This is a consequence of  three possible values of $k_*$ in the $p_k-k$ plane; as a result,  we find a region with  3-DQPT for a given $n$  in the ($\mu_i - \alpha$) plane (with fixed  $\mu_f$) which are also detected by the DTOP.  Interestingly, the width of this region shrinks with increasing $\alpha$ and vanishes in as $\alpha \to 2$ confirming that this unusual region emerges as a result of the long-range interactions.  }
Given the recent experimental realisation of the long-range interacting systems \ct{expt_longrange} and experimental detection of DQPTs \ct{Jurcevic16,flaschner16}, we believe that our studies
can be experimentally verified.

 \begin{acknowledgments}
We acknowledge Utso Bhattacharya for helpful comments and discussions and SERB, DST, New Delhi for financial support.

\end{acknowledgments}



\begin{thebibliography}{11}
\bi{bloch08} I. Bloch, J. Dalibard, and W. Zwerger, Rev. Mod. Phys. {\bf 80}, 885 (2008).
%
\bi{lewenstein12} M. Lewenstein, A. Sanpera, and V. Ahufinger, (Oxford University Press, Oxford (2012)).
%
\bi{greiner02} M. Greiner , O. Mandel, T.  W. Hansch, and  I. Bloch,  Nature {\bf 419}, 51 (2002).
%
%
\bi{kinoshita06} T. Kinoshita, T. Wenger, and D. S. Weiss, Nature {\bf 440}, 900 (2006).
%
\bi{gring12} M. Gring, M. Kuhnert, T. Langen, T. Kitagawa, B. Rauer, M. Schreitl, I. Mazets1, D. Adu Smith, E. Demler, and J. Schmiedmayer, Science {\bf 337}, 1318 (2012).
%
\bi{trotzky12}  S. Trotzky,	Y-A. Chen,	A. Flesch,	I. P. McCulloch,	U. Schollwšck, J. Eisert, and I. Bloch, Nature {\bf 8}, 325 (2012).
%
\bi{cheneau12}  M. Cheneau,	P. Barmettler,	D.  Poletti, M. Endres,	P.  Schauss,	T. Fukuhara,	C. Gross,	I. Bloch,	C.  Kollath, and S.  Kuhr, Nature {\bf 481}, 484  (2012).  
%
\bi{schreiber15}     M.  Schreiber,  S. S. Hodgman, P.  Bordia,  Henrik P. L\" uschen, M. H. Fischer, R. Vosk, E. Altman, U. Schneider, I. Bloch, Science {\bf 349}, 842 (2015). 
%
\bi{fausti11} D. Fausti, R. I. Tobey, , N. Dean,  S. Kaiser, A. Dienst, M. C. Hoffmann, S. Pyon, T. Takayama, H. Takagi,, A. Cavalleri, Science {\bf 331}, 189 (2011). 
%
\bi{rechtsman13}  M. C. Rechtsman,	J. M. Zeuner,	Y.  Plotnik,	 Y.  Lumer,	D.Podolsky,	F.  Dreisow,	S. Nolte,	M. Segev,	and  A. Szameit,    Nature
{\bf 496}  196 (2013).
%
%
\bi{calabrese06} P. Calabrese and  J. Cardy, Phys. Rev. Lett. {\bf 96}, 136801 (2006); J. Stat. Mech,
P06008 (2007).
%
\bi{rigol08} M. Rigol, V. Dunjko, and M. Olshanii, Nature {\bf 452}, 854 (2008).
\bibitem{oka09} T Oka and H Aoki, Phys.  Rev.  B {\bf 79}, 081406 (2009).

%
\bibitem{kitagawa10} T. Kitagawa, E. Berg, M. Rudner, and E. Demler, Phys. Rev. B
{\bf 82}, 235114 (2010).
%
\bibitem{lindner11} N. H. Lindner, G. Refael, and V. Galitski, Nat. Phys. {\bf 7}, 490-495, (2011). 

\bi{bermudez09} A. Bermudez, D. Patane,  L. Amico, and M. A. Martin-Delgado, Phys. Rev. Lett. {\bf 102}, 135702, (2009).

\bi{patel13} A. A. Patel, S. Sharma, and A. Dutta, Eur. Phys. Jour. B {\bf 86}, 367 (2013); A. Rajak and  A. Dutta,  Phys. Rev. E {\bf 89}, 042125 (2014); P. D. Sacramento, Phys. Rev. E {\bf 90} 032138, (2014); M. D. Caio, N. R. Cooper, and M. J. Bhaseen, Phys. Rev. Lett. {\bf 115}, 236403 (2015).    

\bibitem{thakurathi13} M. Thakurathi, A. A. Patel, D. Sen, and A. Dutta, Phys. Rev. B {\bf 88}, 155133 (2013).

\bibitem{mukherjee09}
V. Mukherjee and A. Dutta, J. Stat. Mech. P05005 (2009).
\bibitem{das10}
A. Das, Phys. Rev. B {\bf 82}, 172402 (2010).
\bibitem{Russomanno_PRL12}
{A. Russomanno,  A. Silva,  and G. E. Santoro} , Phys. Rev.
Lett. {\bf 109}, 257201 (2012).

\bi{sharma14}  S. Sharma, A. Russomanno, G. E. Santoro, and A. Dutta, EPL {\bf 106},  67003 (2014).
\bi{bukov16} 
M. Bukov, L. D'Alessio, and A. Polkovnikov, 
Adv. Phys. {\bf 64} , No. 2, 139-226 (2016).
\bibitem{anirban15}A. Dutta, A. Das, and K. Sengupta, Phys. Rev. E,{\bf 110}, 012104 (2015).
\bibitem{arnab16}A. Sen, S. Nandy, and K. Sengupta, Phys. Rev. B,{\bf 94}, 214301 (2016).
%
%
\bibitem{pal10} A. Pal and D. A. Huse, Phys. Rev. B {\bf 82}, 174411  (2010).
\bi{nandkishore15} R. Nandkishore and D. A. Huse, Annual Review of Condensed Matter Physics,  {\bf 6}, 15-38 (2015).  
\bibitem {dziarmaga10} J. Dziarmaga, Advances in Physics  {\bf 59}, 1063 (2010).
\bibitem{polkovnikov11} A. Polkovnikov, K. Sengupta, A. Silva, and M. Vengalattore, \textit{Colloquium: Nonequilibrium dynamics of closed interacting quantum systems}, Rev. Mod. Phys. {\bf 83}, 863 (2011).	
\bi{dutta15} A. Dutta, G. Aeppli, B. K. Chakrabarti, U. Divakaran, T. 
Rosenbaum, and D. Sen, \textit{Quantum Phase Transitions in Transverse Field 
	Spin Models: From Statistical Physics to Quantum Information} (Cambridge 
University Press, Cambridge, 2015).
\bi{eisert15} J. Eisert, M. Friesdorf, and C. Gogolin, Nat. Phys. {\bf 11}, 124 (2015).
\bi{alessio16} L. D'Alessio, Y.  Kafri, A. Polkovnikov, and M. Rigol, Adv. Phys. {\bf 65}, 239 (2016).
%
\bi{jstat} {\it Quantum Integrability in Out of Equilibrium Systems} Edited by  P. Calabrese, F. H. L. Essler, and G. Mussardo,  Special issue of J. Stat. Mech. Th. and Exp. {\bf 2016}, 
(2016).
%
%
\bi{expt_longrange}P.  Richerme, Z.-X. Gong, A. Lee, C. Senko, J. Smith, M. Foss-Feig, S. Michalakis, A. V. Gorshkov, and C. Monroe, Nature {\bf 511}, 198 (2014).
%
\bi{silva_longrange}B \v Zunkovi\v c, A. Silva, and M. Fabrizio, Phil. Trans. R. Soc. A {\bf 374}, 20150160 (2016).
%
\bi{vodola}D. Vodola, L. Lepori, E. Ercolessi, and G. Pupillo, New J. Phys. {\bf 18}, 015001 (2016).
\bi{vodola1}O. Viyuela, D. Vodola, G. Pupillo, and M. A. Martin-Delgado, Phys. Rev. B {\bf 94}, 125121 (2016).
\bi{vodola2}L. Lepori, A. Trombettoni, and D. Vodola, J. Stat. Mech. 033102 (2017).
\bi{regemortel}M. V. Regemortel, M. Wouters, and D. Sels,Phys. Rev. A {\bf 93}, 032311 (2016).
\bi{unpub}S. Nandy, A. Sen, and K. Sengupta, unpublished,(2017).
\bi{dellanna17}Antonio Alecce, Luca Dell'Anna, arXiv:1703.10086 (2017).
\bi{fey2016} S. Fey and K. P. Schmidt, Phys. Rev. B {\bf 94}, 075156 (2016).

\bi{spin_model_dpt}B \v Zunkovi\v c, M. Heyl, M. Knap, A. Silva, arXiv:1609.08482 (2016)
\bi{halimeh17} J. C. Halimeh, V. Zauner-Stauber, I. P. McCulloch, I. de Vega, U. Schollwšck, and M. Kastner, Phys. Rev. B {\bf 95}, 024302 (2017).

\bi{halimeh171} J. C. Halimeh and V. Zauner-Stauber, arXiv:1610.02019  (2017).

\bi{homrighausen17} I. Homrighausen, N. O. Abeling, V. Zauner-Stauber, and J. C. Halimeh, arXiv:1703.0919 (2017).

\bi{kzml} D. Jaschke, K. Maeda, J. D. Whalen, M. L. Wall, and L. D. Carr,
New J. Phys. {\bf 19} , 033032 (2017).

\bi{ruelles68} D. Ruelle, Commun. Math. Phys. {\bf 9}, 267 (1968).
\bi{dyson69}F. J. Dyson, Commun. math. Phys. {\bf 12}, 91?107 (1969).
\bi{dysonn69}F. J. Dyson, Commun. math. Phys. {\bf 12}, 212-215 (1969).
\bi{kac69}M. Kac, and C. J. Thompsom, J. Math. Phys. {\bf 10}, 8 (1969).
\bi{thouless69} D. J. Thouless, Phys. Rev. {\bf 187}, 732 (1969).


\bi{fisher72} M.E. Fisher, S.K. Ma, and B.G. Nickel, Phys. Rev. Lett. {\bf 29}, 917 (1972). 

\bi{yuval71} P.W. Anderson and G. Yuval, J. Phys. C {\bf 4}, 607 (1971).


\bi{kosterlitz76} J.M. Kosterlitz, Phys. Rev. Lett. {\bf 37}, 1577 (1976).

\bi{cardy81}J.L. Cardy, J. Phys. A {\bf 14}, 1407 (1981).
%
%
\bi{bhattacharjee81} J. Bhattacharjee, S. Chakravarty, J.L. Richardson, and D.J. Scalapino, Phys. Rev. B {\bf 24}, 3862 (1981).
\bi{bhattacharjee82} J.K. Bhattacharjee, J.L. Cardy, and D.J. Scalapino, Phys. Rev. B {\bf 25}, 1681 (1982)
\bi{imbrie88} J.Z. Imbrie and C.M. Newmann, Commun. Math. Phys. {\bf 118}, 303 (1988).
\bi{luijten01} E. Luijten and H. Me§ingfeld, Phys. Rev. Lett. {\bf 86}, 5305 (2001).

\bibitem{sachdev96} S. Sachdev, \textit{Quantum Phase Transitions}
(Cambridge University Press, Cambridge, UK, 2011).
%
\bibitem{suzuki13} S. Suzuki, J-i Inoue, and Bikas K. Chkarabarti,  \textit{Quantum Ising Phases and Transitions in Transverse Ising Models} (Springer, Lecture Notes in Physics, Vol. 862 (2013)).


\bi{dutta01} A. Dutta and J. K. Bhattacharjee, Phys. Rev. B {\bf 64}, 184106 (2001).

\bi{kitaev01} A. Kiteav, arXiv:cond-mat/0010440 (2000); A.  Kitaev, C.  Laumann, arXiv:0904.2771, Les Houches Summer School "Exact methods in low-dimensional physics and quantum computing" , 2008.


\bibitem{Kibble76} T. W. B. Kibble,
{J. Phys. A: Math. Gen.  {\bf 9}, 1387 (1976)};
{Phys. Rep. {\bf 67,} 183 (1980).}

\bibitem{Zurek96} W. H. Zurek,
{Nature (London) {\bf 317}, 505 (1985)};
{Acta Phys. Pol. B {\bf 24}, 1301 (1993)}; 
{Phys. Rep. {\bf 276}, 177 (1996).}
%

\bibitem{Damski05}\  B. Damski,
{Phys. Rev. Lett. {\bf 95}, 035701 (2005).}
%
\bibitem{Zurek05} \  W.H. Zurek,
{Phys. Rev. Lett. {\bf 95}, 105701 (2005).}
%
\bibitem{ZDZ05}\  W. H. Zurek, U. Dorner,  and P. Zoller, 
{Phys. Rev. Lett. {\bf 95}, 105701 (2005).}
%
\bibitem{Polkovnikov05}\  A. Polkovnikov, 
{Phys. Rev. B {\bf 72}, 161201(R) (2005).}
%
\bibitem{damski_zurek06}\  B. Damski and W. H. Zurek, 
{Phys. Rev. A {\bf 73}, 063405 (2005).}
\bibitem{mukherjee07} \ V. Mukherjee, U. Divakaran, A. Dutta, and D. Sen
{Phys. Rev. B {\bf 76}, 174303 (2007).}
\bibitem{divakaran08} \ U. Divakaran, A. Dutta, and D. Sen
{Phys. Rev. B {\bf 78}, 144301 (2008).}
\bibitem{shreyoshi08} \ K. Sengupta, D. Sen, and S. Mondal
{Phys. Rev. Lett. {\bf 100}, 077204 (2008).}
\bibitem{sen08} \ D. Sen, K. Sengupta, and S. Mondal
{Phys. Rev. Lett. {\bf 101}, 016806 (2008).}
\bibitem{anirban16} \ A. Dutta,A. Rahmani, and A. del Campo
{Phys. Rev. Lett. {\bf 117}, 080402 (2016).}
%
\bi{heyl13} M. Heyl, A. Polkovnikov, and S. Kehrein, Phys. Rev. Lett., {\bf 110}, 135704 (2013).

\bi{karrasch13} C. Karrasch and D. Schuricht,  Phys. Rev. B, {\bf 87}, 195104 (2013).

\bi{kriel14} N. Kriel, C. Karrasch, and S. Kehrein, Phys. Rev. B {\bf 90}, 125106 (2014).

\bi{heyl14} M. Heyl, Phys. Rev. Lett., {\bf 113}, 205701 (2014).

\bi{pollmann10} F. Pollmann, S. Mukerjee, A. G. Green, and J. E. Moore, Phys. Rev. E {\bf 81}, 020101(R) (2010).

\bi{divakaran16} U. Divakaran, S. Sharma, and A. Dutta, Phys. Rev. E {\bf 93}, 052133 (2016).
%
\bi{landau} C. Zener, Proc. Roy. Soc. London Ser A {\bf 137}, 696 (1932); L. D.
Landau and E. M. Lifshitz, {\it Quantum Mechanics: Non-relativistic Theory}, 
2nd ed. (Pergamon Press, Oxford, 1965).

\bi{sei} S. Suzuki and M. Okada, in {\it Quantum Annealing and Related 
	Optimization Methods}, Ed. by A. Das and B. K. Chakrabarti (Springer-Verlag,
Berlin, 2005), p. 185.

\bi{vitanov} N. V. Vitanov and B. M. Garraway,  Phys. Rev. A {\bf 53}, 4288 (1996); N. V. Vitanov, {\it ibid.} {\bf 59}, 988 (1999).
%
\bi{functions} F. W. J. Olver, D. W. Lozier, R. F. Boisvert, and C. W. Clark, \textit{NIST Handbook of Mathematical Functions}
(Cambridge University Press, Cambridge, England, 2010); M. Abramowitz and I. A. Stegun, \textit{Handbook of Mathematical Functions}(Dover, 1964).
%
\bi {fisher65} M.E. Fisher, in {\it Boulder Lectures in Theoretical Physics} (University of Colorado, Boulder, 1965), Vol. 7.
\bi {lee52} C. Yang and T. Lee, Phys. Rev. {\bf 87}, 404 (1952).
\bi{saarloos84} W. van Saarloos and D. Kurtze, J. Phys. A {\bf 17}, 1301 (1984).
%
\bi{heyl15} M. Heyl, Phys. Rev. Lett., {\bf 115}, 140602 (2015) .

\bi{palami15} T. Palmai, Phys. Rev. B {\bf 92}, 235433 (2015).

\bi{huang16} Z.  Huang and A.  V. Balatsky, Phys. Rev. Lett. {\bf 117}, 086802 (2016).

\bi{puskarov16} T. Puskarov and D. Schuricht, SciPost Phys. {\bf 1}, 003 (2016).

\bi{zhang16} J. M. Zhang and  H.-T. Yang, EPL {\bf 116}, 10008 (2016).

\bi{heyl16} M. Heyl, Phys. Rev. B {\bf 95}, 060504 (2017).

\bi{vajna14} S. Vajna, and B. Dora,  Phys. Rev. B {\bf 89}, 161105(R) (2014).

\bi{sharma15} S. Sharma, S. Suzuki, and A. Dutta,  Phys. Rev. B {\bf 92}, 104306 (2015).

\bi{vajna15} S. Vajna and B. Dora, Phys. Rev. B {\bf 91}, 155127 (2015).

\bi{schmitt15} M. Schmitt and S. Kehrein, Phys. Rev. B {\bf 92}, 075114 (2015).

\bi{budich15} J. C. Budich and  M. Heyl,  Phys. Rev. B {\bf 93}, 085416 (2016). 
%
\bi{andraschko14} F. Andraschko and J. Sirker, Phys. Rev. B {\bf 89}, 125120 (2014).

\bi{canovi14} E. Canovi, P. Werner, and M. Eckstein, Phys. Rev. Lett. {\bf 113}, 265702 (2014).

\bi{sharma16} S. Sharma, U. Divakaran, A. Polkovnikov, and A. Dutta, Phys. Rev. B {\bf 93}, 144306 (2016).

\bi{zvyagin17} A.A. Zvyagin, Low Temp. Phys. {\bf 42}, 971-994 (2016).

\bi{sei17} T.  Obuchi, S. Suzuki, K. Takahashi, arXiv:1702.05396 (2017).

\bi{fogarty17} Thoms Fogarty, Ayaka Usui, Thomas Busch, Alessandro Silva,
John Goold,  arXiv:1704.07659 (2017).

\bi{utso16} U. Bhattacharya and A. Dutta, arXiv:1610.02674 (2016).

\bi{utso17} U. Bhattacharya and A. Dutta, arXiv:1701.03911 (2017).

\bi{bhattacharya17} U. Bhattacharya, S. Bandyopadhyay and A. Dutta, arXiv:1705.04555  (2017).

\bi{heyl17} M. Heyl and J.  C. Budich, arXiv:1705.08980  (2017).
%



%
 \bibitem{Jurcevic16}  P. Jurcevic, H. Shen, P. Hauke, C. Maier, T. Brydges, C. Hempel, B. P. Lanyon, M. Heyl, R. Blatt, C. F. Roos,  arXiv:1612.06902 (2016).
%
\bi{flaschner16} N. Fl\"aschner, D. Vogel, M. Tarnowski, B, S. Rem, D.-S. L\"uhmann, M. Heyl, J.  Budich, L. Mathey, K. Sengstock, C. Weitenberg,
arXiv:1608.05616 (2016).
%



%
\end{thebibliography}
\end{document}